\documentclass{article}

\usepackage{arxiv}

\usepackage[utf8]{inputenc} 
\usepackage[T1]{fontenc}    
\usepackage{hyperref}       
\usepackage{url}            
\usepackage{booktabs}       
\usepackage{amsfonts}       
\usepackage{nicefrac}       
\usepackage{microtype}      
\usepackage{lipsum}
\usepackage{graphicx}

\usepackage{amssymb}
\usepackage{color}
\usepackage{xcolor}
\usepackage{multirow}
\usepackage{colortbl}
\usepackage{mathrsfs}
\usepackage{amsmath}
\usepackage{tabularx}
\usepackage{textcomp,mathcomp}
\graphicspath{ {./images/} }

\title{Accelerated Proton Resonance Frequency-based Magnetic Resonance Thermometry by Optimized Deep Learning Method}

\author{
 Sijie Xu$^{1,\dag}$ \\
  \texttt{sijie.x@sjtu.edu} \\
   \And
 Shenyan Zong$^{2,\dag}$ \\
  \texttt{shenyanzong@fudan.edu.cn} \\
  \And
 Chang-Sheng Mei$^{3,4}$ \\
  \texttt{mei.changsheng@gmail.com} \\
  \And
  Guofeng Shen$^{1,*}$ \\
  \texttt{shenguofeng@sjtu.edu.cn} \\
  \And
  Yueran Zhao$^{1}$ \\
  \texttt{shenguofeng@sjtu.edu.cn} \\
  \And
  He Wang$^{2,5,*}$ \\
  \texttt{shenguofeng@sjtu.edu.cn} \\
}

\begin{document}

\maketitle

{\fontsize{8pt}{10pt}\selectfont $^1$Biomedical Instrument Institute, School of Biomedical Engineering, Shanghai Jiao Tong University, Shanghai.\\}
{\fontsize{8pt}{10pt}\selectfont $^2$Institute of Science and Technology for Brain-Inspired Intelligence, Fudan University, Shanghai.\\}
{\fontsize{8pt}{10pt}\selectfont $^3$Department of Radiology, Brigham and Women’s Hospital, Harvard Medical School, Boston, Massachu-setts.\\}
{\fontsize{8pt}{10pt}\selectfont $^4$Department of Physics, Soochow University, Taipei\\}
{\fontsize{8pt}{10pt}\selectfont $^5$Department of Radiology, Shanghai Fourth People’s Hospital Affiliated to Tongji University School of Medicine, Shanghai\\}
{\fontsize{8pt}{10pt}\selectfont $^\dag$Co-First Author, $^*$Corresponding Author\\}

\begin{abstract}
\textbf{Background:} Proton resonance frequency (PRF)--based magnetic resonance (MR) thermometry is essential in thermal ablation therapies through focused ultrasound (FUS). The clinical treatments require temperature feedback must be rapid and accurate.
Purpose: This work aims to enhance temporal resolution in dynamic MR temperature map reconstruction with an improved deep learning method, to ensure the safety and effectiveness of FUS treatments.

\textbf{Methods:} The training-optimized methods and five classical neural networks were applied on the 2-fold and 4-fold under-sampling k-space data to reconstruct the temperature maps. The used neural networks were cascade net, complex valued U-Net, shift window transformer for MRI, real valued U-Net and U-Net with residual block. The enhanced training modules included offline/online data augmentations, knowledge distillation, and the amplitude-phase decoupling loss function. The heating experiments were performed by a FUS transducer on \textit{phantom} and \textit{ex vivo} tissues, respectively. In datasets, the ground-truth was the complex MR images with accurate temperature increases. These data were also manually under-sampled to imitate acceleration procedures and trained in our method to get the reconstruction model. The additional dozen or so testing datasets were separately obtained for evaluating the real-time performance and temperature accuracy.

\textbf{Results:} Acceleration factors of 1.9 and 3.7 were found for $2\times$ and $4\times$ k-space under-sampling strategies and the ResUNet-based deep learning reconstruction performed exceptionally well. In 2-fold acceleration scenario, the RMSE of temperature map patches provided the values of 0.888 ℃ and 1.145 ℃ on \textit{phantom} and \textit{ex vivo} testing datasets. The DICE value of temperature areas enclosed by $43$ ℃ isotherm was 0.809, and the Bland-Altman analysis showed a bias of $-0.253$ ℃ with the apart of $\pm2.16$ ℃. In $4\times$ under-sampling case, these evaluating values decreased by approximately 10\%.

\textbf{Conclusion:} This study demonstrates that the application of deep learning-based reconstruction significantly enhances the accuracy and efficiency of MR thermometry, particularly benefiting the clinical thermal therapies for uterine fibroid, essential tremor, and prostate cancer by FUS.

The source code for our optimizing methods and neural networks is available at: \url{https://github.com/minipuding/FastMRT}.

\end{abstract}


\section{Introduction}
Magnetic resonance (MR) thermometry is widely used in noninvasive surgical treatments by focused ultrasound (FUS)\cite{hynynen2010mri}. However, achieving real-time temperature measurement that preserves temperature information using magnetic resonance imaging (MRI) is highly challenging due to the underlying imaging principles. Such surgeries require the continuous acquisition of approximately ten frames for each ablation, with each frame taking approximately two seconds. A complete uterine fibroid ablation procedure typically lasts between two and four hours, with a considerable portion of the time dedicated to image acquisition. The use of temperature monitoring intervals lasting seconds still poses a certain degree of safety risk to patients\cite{yuan2012towards}. Throughout the procedure, patients must remain motionless to prevent any hazardous circumstances, which can be highly distressing for them.

Fast imaging based on the reconstruction of under-sampled magnetic resonance images has been in development for several years and has shown great potential to significantly increase the speed of temperature measurement. Most studies in the area of MRI temperature measurement use conventional methods, such as parallel imaging and compress sensing\cite{cao2019low, shimronTemporalDifferencesTED2020}. The accelerated MR thermometry through coil-sensitivity encoding was generally accompanied by a decrease in temperature accuracy\cite{meiCombiningTwodimensionalSpatially2011}. Besides that, the reduced field of view (FOV) method might be an alternative for fast temperature measurements, but the absence of a full-FOV monitor increased the risk of ablation treatments. More advanced non-cartesian readout strategies such as spiral and radial MR thermometry proposed by Kisoo and Pooja et al, can achieve volumetric and motion-immune temperature measurements with a temporal resolution of 100~300 ms for each slice\cite{gaurAcceleratedMRIThermometry2015,kim2023motion}. However, our novel deep learning-based rapid reconstruction method was non-conflicting with these sampling strategies. The spiral and radial k-space data can also be under-sampled to reach higher imaging speeds, and accelerated imaging via reconstruction in this study was also applicable. The work to be done is to train a specialized reconstruction model for them through our proposed method. In addition, the rapid echo planar imaging (EPI) sequence was validated for temperature measurements by Andrew and Henrik et al\cite{odeen2014sampling,holbrook2010real}. Nevertheless, the segmented or single-shot EPI was always vulnerable to B0 inhomogeneities. The irresistible susceptibility led to a significant reduction in the clinical acceptability of using this sequence for temperature measurements\cite{zong2020improved,mei2015accurate}. Furthermore, the temperature increase-induced focus shift remains a difficult and not well-resolved problem. In contrast, the rapid reconstruction method we proposed for cartesian-based gradient echo sequences was more robust.

Since the inception of the fast MRI challenge\cite{zbontarFastMRIOpenDataset2019,knollFastMRIPubliclyAvailable2020}, numerous deep learning-based MRI reconstruction techniques\cite{muckleyResults2020FastMRI2021,palReviewExperimentalEvaluation2022} have been proposed. Following the emergence of Vision Transformer (ViT)\cite{dosovitskiyImageWorth16x162021}, several transformer-based reconstruction methods have also been proposed\cite{linVisionTransformersEnable2022,guoReconFormerAcceleratedMRI2022,huangSwinTransformerFast2022}. However, magnetic resonance thermometry encounters two major issues when attempting to apply existing fast imaging algorithms. Firstly, the most commonly used temperature measurement method, proton resonance frequency (PRF) shift\cite{poorterNoninvasiveMRIThermometry1995}, relies on the phase discrepancy of complex images. However, current fast MRI methods primarily focus on reconstructing amplitude images, with little emphasis on phase\cite{jong-minRealTimeT1PRFBased2019}. As a result, these methods are suboptimal for preserving phase information in magnetic resonance temperature measurement, lacking specific design or improvement for this purpose. Secondly, current undersampling reconstruction methods prioritize image quality restoration over time preservation, leading to increasingly larger and more complex models with insufficient attention to the impact of model inference time on actual acceleration rates\cite{shamshadTransformersMedicalImaging2022}. Furthermore, insufficient datasets may result in limited model's performance due to overfitting and underutilization.

This work improves the performance of deep learning by adopting network structure-independent methods without increasing the number of network parameters or computational complexity to achieve fast and accurate measurement. The proposed approach involves several techniques to improve the performance of neural network models; it utilizes offline diffusion model augmentation, online complex-valued data augmentation techniques, knowledge distillation, and an amplitude-phase decoupled loss function. The first two modules are utilized for data augmentation to prevent overfitting and unleash the potential of the model. The knowledge distillation module enables a smaller model to learn capabilities several times greater than its parameter capacity. The decoupled loss function separates amplitude and phase differences, allowing the model to adjust weights and focus on the image phase. Based on these training strategies, the cascade net (CasNet)\cite{schlemperDeepCascadeConvolutional2017}, complex valued U-Net (CUNet)\cite{dedmariComplexFullyConvolutional2018}, shift window transformer for MRI (SwinMR)\cite{huangSwinTransformerFast2022,ekanayakeMultiheadCascadedSwin2022,liuSwinTransformerHierarchical2021}, real valued U-Net (RUNet)\cite{ronnebergerUNetConvolutionalNetworks2015} and U-Net with residual block (ResUNet) were involved in this deep learning method to improve the speed of MR temperature measurements.

\section{Theories}
\label{sec:headings}
\subsection{MR Reconstruction via Deep Learning}
The speed of MRI is determined by the number of sampled k-space lines for a regular gradient echo sequence, and the acceleration can be achieved through reducing phase encoding numbers. In the case of single-channel MRI signal sampling, it can be mathematically represented by the formula \ref{(1)}:

\begin{equation}
y=\mathcal{M}\cdot \mathscr{F}(x)+\epsilon
\label{(1)}
\end{equation}

where $x \in C^{(N_1\times N_2)}$  denotes the MR images reconstructed from fully sampled k-space, while $\mathcal{M} \in C^{(N_1\times N_2)}$ represents the mask lines selected from the phase encoding direction of $y \in C^{(N_1\times N_2)}$ , and $\mathscr{F}(\cdot)$ is the Fourier transform.

A typical function for estimating the MR image $x$ from measurements is given by:
\begin{equation}
x=\mathop{\arg\min}\limits_{x}||y-\mathcal{M}\cdot \mathscr{F}(x)||_2^2 + \lambda\cdot R(x)
\end{equation}

where $R(x)$ denotes the regularizer, which is dependent on the reconstruction algorithm used.
For deep learning training processes, a function $G_{DL}(\cdot)$ can be utilized as the regularizer:
\begin{equation}
    \begin{split}
        \hat{x} = arg\min\limits_{x}||y-\mathcal{M}\cdot\mathscr{F}(x)||_2^2+\lambda R(x,\theta^*), \\
        where\ \theta^* = arg\min\limits_{\theta}E||x - G_{DL}(\mathcal{M}\cdot\mathscr{F}(x);\theta)||, x\sim S
    \end{split}
\end{equation}

where $S$ is the dataset and $x$ is the complex image sampled from $S$. We train a model to minimize the expected difference between sampled and fully sampled images\cite{palReviewExperimentalEvaluation2022}.

\subsection{Proton Resonance Frequency Shift}

At present, proton resonance frequency (PRF) shift thermometry is a widely used technique for temperature measurement via MRI. PRF shift thermometry shows a persistent linear correlation with temperature and is largely tissue-type agnostic (excluding adipose tissue), while providing a simple and robust real-time measurement method through the regular sequences. To determine temperature changes it calculates the phase difference between the magnetic resonance images with heating and the baseline images. The temperature alteration can be expressed as a linear function of the phase difference, as shown by formula \ref{(4)}:

\begin{equation}
    \Delta T=\frac{\phi-\phi_{ref}}{\alpha\cdot\gamma\cdot  t_{TE}\cdot B_0}
\label{(4)}
\end{equation}

where $\phi$ represents the phase of the current image, $\phi_{ref}$ represents the phase of the image acquired at time 0, $\alpha$ denotes the PRF (Proton Resonance Frequency) change coefficient of water tissue, which is -0.01 ppm/℃, $\gamma$ represents the magnetic moment ratio of hydrogen atoms, $t_{TE}$ denotes the echo time, and $B_0$ represents the main magnetic field strength. 

\subsection{Actual Acceleration Ration}
As stated above, it was assumed that the under-sampling rate represented the time saved, without considering the inference time required by the reconstruction algorithm itself. However, for magnetic resonance temperature measurement tasks, we need real-time imaging. Therefore, it is essential to compute the effective acceleration rate of the reconstruction model, which is defined as :
reconstruction model. We define it as follows:
\begin{equation}
    E_{N=n}=\frac{t_a}{\frac{t_a}{n}+t_m}=\frac{n\cdot t_a}{t_a+n\cdot t_m}
    \label{(5)}
\end{equation}

where $t_a$ denotes the acquisition time of the fully sampled image, $t_m$, $E_{N=n}$ and $n$ represent the inference time of the model, the effective acceleration rate, and the theoretical acceleration rate, respectively. When computing this metric in practice, we approximate $t_a$ by $t_{TR}\times num_{pe}$ where the $num_{pe}$ denotes the number of phase encoding and we obtain $t_m$ via the model’s CPU forward inference time.

\section{Methods}
\subsection{Deep Learning Training and Models}

\begin{figure*}[t]
\centerline{\includegraphics[width = 1.0\textwidth]{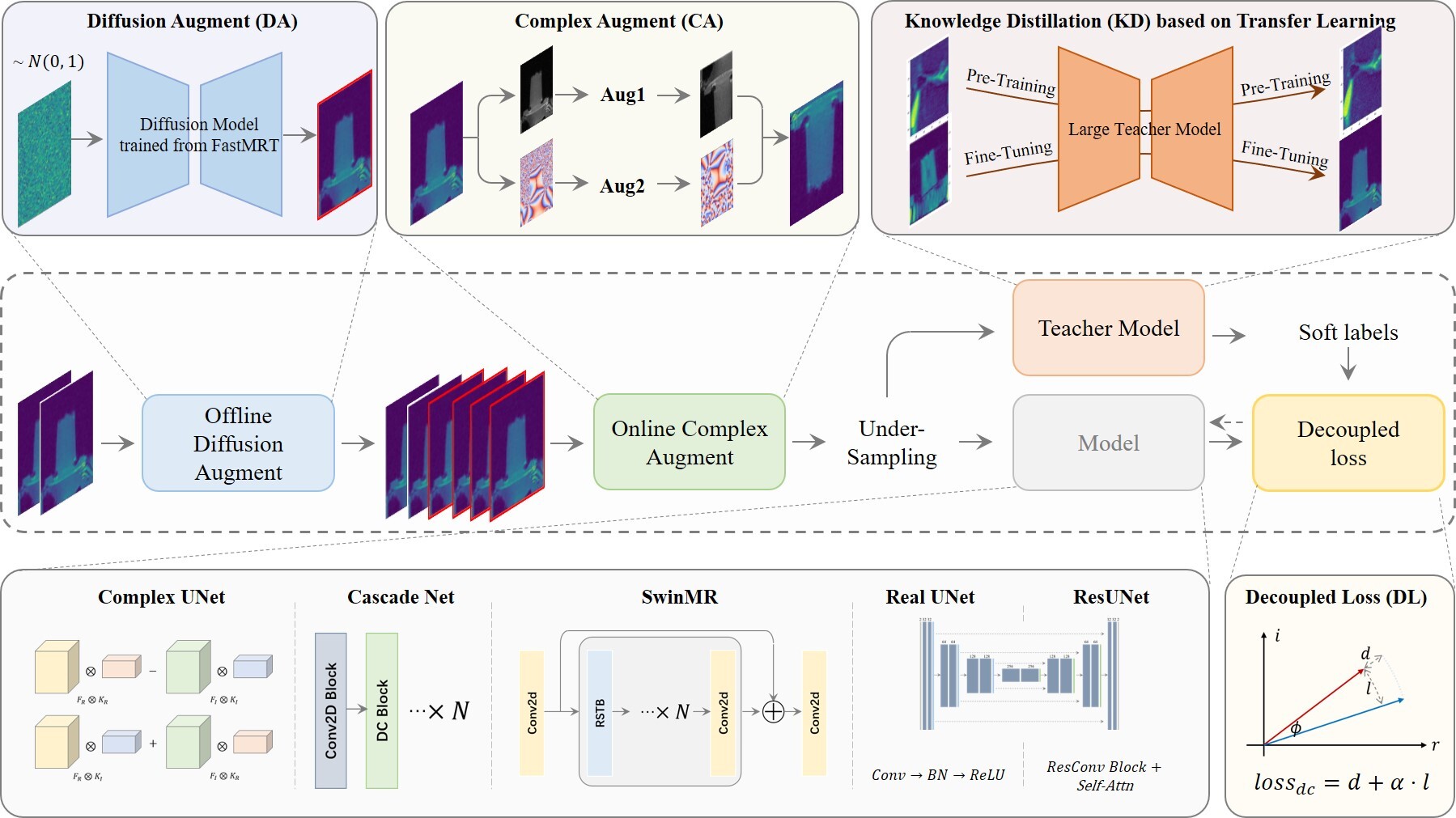}}
\caption{Algorithm Structure Diagram showcasing four optimizing modules: {\textcolor[rgb]{0.25, 0.5, 0.75}{\textbf{offline diffusion augments (DA)}}}, {\textcolor[rgb]{0.25, 0.75, 0.5}{\textbf{online complex augments (CA)}}}, {\textcolor[rgb]{0.75, 0.5, 0.20}{\textbf{knowledge distillation (KD)}}}, and {\textcolor[rgb]{0.8, 0.8, 0.25}{\textbf{decoupled loss (DL)}}}. The DA module employs a trained diffusion model to generate new data samples offline, thereby enhancing data diversity and complexity. The CA module combines augmented amplitude and phase maps of complex data. The KD module extracts knowledge from a larger pretrained teacher model and transfers it to a smaller model, thereby enhancing performance using a compact model. The teacher model is pretrained from FastMRI dataset and fine-tuned on our dataset. The DL module separates the amplitude and phase components of a signal, assigning distinct weights to each, to enhance the reconstruction capability of the phase component. The study incorporates five classical models.}
\label{framework}
\end{figure*}

As depicted in Figure \ref{framework}, the proposed deep learning method incorporates data augments, teacher model, decoupled loss function, and classical network models. The offline diffusion augment and online data augment were able to expand the amount of MR images obtained in the heating experiments. This preprocessing procedure was designed to improve the performance of network model. Furthermore, the cascade net, complex net, swin-transformer, real-unet, and resunet were used in the training to get five different reconstruction models. The decoupling loss function was modified to adapt complex-valued MR images used for temperature measurements. During the training process, we leverage a sizable teacher model pre-trained on the FastMRI dataset and further fine-tuned on our own dataset. This teacher model serves as a guiding influence for the student model, enabling us to achieve a compact model with a performance comparable to that of the teacher model.
\subsubsection{Offline Diffusion Augmentations}
In medical image-related tasks, there is a growing trend toward the adoption of model-based data augmentation techniques\cite{kebailiDeepLearningApproaches2023}. As demonstrated by Trabucco Brandon et al, the diffusion model is a more effective means of generating diverse and realistic images\cite{trabuccoEffectiveDataAugmentation2023}. Therefore, the diffusion model called Denoising Diffusion Probabilistic Model (DDPM)\cite{hoDenoisingDiffusionProbabilistic2020} was utilized here to generate a substantial amount of similar data for our training process. Additionally, we set the time step to 600 and increased the amount of data by a factor of five for the \textit{phantom} and \textit{ex vivo} sub-datasets, respectively. Since the diffusion model requires a considerable amount of time for processing, we opted for an offline method to generate augmented data before the training phase.
\subsubsection{Online Complex Augmentations}
Before inputting data into the model, we perform conventional data augmentation, which involves random cropping, flipping, rotation (0°, 90°, 180°, 270°), and Gaussian blurring. We would like to high-light that we extended real-valued data augmentation to complex-valued data by separately augmenting the magnitude and phase of complex images before combining them. This approach significantly increases data diversity by orders of squares, thereby enhancing the model's robustness and generalizability. To avoid introducing undesirable bias into the model training due to the disruption of spatial consistency between magnitude and phase, we apply complex-valued data augmentation with a specific probability. We set the optimal probability for triggering complex-valued data augmentation to 0.3.
\subsubsection{Base Models}
To achieve faster temperature map reconstruction, we implemented our proposed method on the Naive-Real-UNet (RUNet) and ResUNet, which has state-of-the-art (SOTA) performance and is lightweight, making it an ideal choice for MR temperature map reconstruction. Compared to RUNet, ResUNet replaces ordinary convolutional layers with residual blocks and adds self-attention modules, which can be considered an improved version of RUNet. We also compared our method with several SOTA MR reconstruction methods, such as the cascade network with data consistency (CasNet)\cite{schlemperDeepCascadeConvolutional2017}, complex-valued convolutional network with UNet structure (CUNet)\cite{dedmariComplexFullyConvolutional2018,trabelsiDeepComplexNetworks2018}, Swin-Transformer (SwinMR)\cite{huangSwinTransformerFast2022}, and the RUNet and ResUNet without any structure modifications.
\subsubsection{Knowledge Distillation}
Knowledge distillation\cite{hintonDistillingKnowledgeNeural2015} is a technique that can accelerate inference times by transferring the knowledge learned by a larger and more complex model to a smaller and simpler one while preserving or even improving performance\cite{murugesanKDMRIKnowledgeDistillation2020,tanSemisupervisedDistillationLearning2022}. Larger models usually exhibit stronger generalization capabilities, but their inference speeds may be lower. Thus, the knowledge acquired by the larger “teacher” model network is transferred to the smaller “student” model in the form of soft labels. Initially, we pre-train a teacher model with an identical structure to that of the student model and extend the channels by a factor of four using both online and offline augmentation techniques. To ensure comprehensive training of the teacher network, we adopt a two-step approach. We perform pre-training on the FastMRI dataset, followed by full parameter fine-tuning on our research dataset. During each forward pass of student model, the output is compared to both the ground truth and the soft labels generated by the pre-trained teacher model. The losses are then weighted and calculated accordingly. The weights are dynamically adjusted to decay over time during training so that the student model can primarily learn from the teacher network in the initial stages and gradually transition to learning from the ground truth. The loss function for knowledge distillation is:
\begin{equation}
    \begin{split}
        L_{total} = (1-w)\cdot L_{gt}+w\cdot L_{soft},  \\
        where\ w=(1-\frac{E_{curr}}{E_{total}})\cdot \gamma
    \end{split}
    \label{fkd}
\end{equation}
where $l_{gt}$ and $l_{soft}$ denote the loss values calculated with the ground truth and the soft labels generated by the teacher network, respectively; $E_{curr}$ and $E_{total}$ denote the current epoch and the total number of epochs, respectively; $\gamma$ is the only hyperparameter that adjusts the weight of the teacher network guidance.

\subsubsection{Decoupled Loss}
\begin{figure}[ht]
\centerline{\includegraphics[width = 0.7\textwidth]{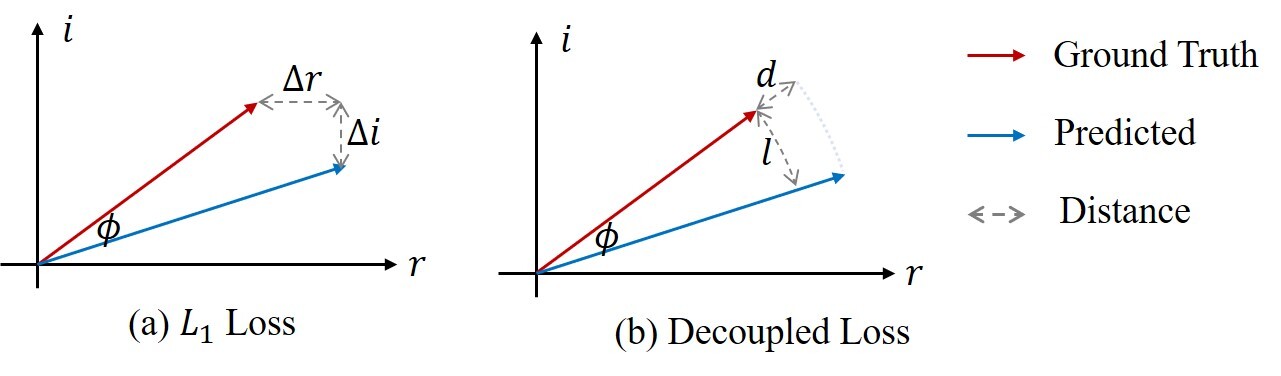}}
\caption{Schematic diagram of the complex loss function. (a) represents the L1 loss function, which is the sum of the differences in the real part $\Delta x$ and imaginary part $\Delta y$; (b) represents the decoupling loss, which is composed of the absolute error $d$ in magnitude and the difference $l$ in phase (angle). The figure provides an intuitive visualization of the components of the complex loss function.}
\label{dl}
\end{figure}

Our investigation revealed that most reconstruction algorithms rely on the L1 loss function\cite{coleAnalysisDeepComplexvalued2021,leeDeepResidualLearning2018} which may not be optimal for temperature measurement tasks that emphasize phases. More specifically, the L1 loss function tends to couple the magnitude and phase, resulting in identical loss values that correspond to different phases. Consequently, it becomes challenging to specifically optimize the phase component.

Therefore, we attempted to decouple the loss function, as illustrated in Figure \ref{dl}. We partitioned the loss into two components: magnitude loss (computed as the absolute error of the amplitude values) and phase loss (computed as the error in radians), with the former quantifying the difference in magnitude and the latter quantifying the phase difference. In contrast to the decoupled loss functions proposed by Zhang et al.\cite{zhangWeightedMagnitudePhaseLoss2021} and other researchers, our decoupled loss function is simpler and has a more straightforward geometric interpretation:

\begin{equation}
    loss_{dc}=d+\alpha\cdot l=||\hat y| - |y||+\alpha\cdot |\hat y|\cdot \mathcal{A}(\hat y\times \bar y)
\end{equation}

where $\mathcal{A}$ represents the angle calculation function and $\alpha$ is a parameter that controls the degree of bias applied to the phase loss. The variables $y$,$\hat y$, and $\bar y$ represent the predicted complex output, ground truth, and the complex conjugate of $y$, respectively. In some other fields related to phase maps, different types of specialized phase loss functions are used\cite{zhangWeightedMagnitudePhaseLoss2021,spoorthiPhaseNetPhaseUnwrapping2020}, but the rationale for their design and mathematical basis are not provided.

\subsection{Heating Experiments and Implementation}
\subsubsection{FUS heating}
Our dataset was acquired through the application of a 128-element high-intensity focused ultrasound transducer (with a frequency of $1.1 MHz$, a focal length of $150 mm$, and a focal radius of $120 mm$), followed by imaging with a 3T MR system (Discovery MR750; GE Healthcare, Milwaukee, WI). 
The images were obtained using the Fast Spoiled Gradient Echo (FSPGR) sequence, with 96 phase encoding steps, a TR/TE of $12/16 ms$, a flip angle of $30^\circ$, a slice thickness of $3 mm$, a field of view (FOV) of $28\times28 cm^2$, a Number of Excitation (NEX) of 1, and a bandwidth of $\pm62.5kHz$. 
The dataset comprises two distinct parts: \textit{phantom} heating data and \textit{ex vivo} heating data. For each part, there are 96 heating samples (consisting of 2186 slices) and 105 samples (consisting of 1623 slices), respectively, with each sample containing either one or three layers. The temperature change at the focus was approximately 30 degrees Celsius, and the focus position was consistently located at the center of the image. 
To enhance the speed of temperature measurement, we employed a smaller TR and fewer phase encoding steps, which led to a lower signal-to-noise ratio and resolution of the acquired images. This underscores the significance of utilizing fast temperature measurement algorithms to compensate for the reduced image quality.
\subsubsection{Model Metrics}
\paragraph{Temperature Metrics.} After deriving the PRF temperature map from the complex images generated by the model and the reference images, we obtain a common metric that characterizes the reconstruction error of the entire image by calculating the average pixel-wise error compared to the original temperature map (represented as $T_{err}$). However, as we use the HIFU device to focus heat on a very small area, only a small region undergoes significant temperature changes. Therefore, it is also necessary to consider local metrics for the heating focus.

Specifically, we evaluate the temperature using metrics such as root mean square error (RMSE), standard deviation (STD), and Dice coefficient (DICE). These metrics are calculated within a pixel block that is cropped around the focal area, covering one-fourth of the width and height of the image. Furthermore, we also assess the agreement between the reconstructed temperature values and the reference values using Bland-Altman analysis and examine the linear relationship between these two sets of values using linear regression analysis, both of which are commonly used in previous related works\cite{zong2020improved}. The temperature map is calculated from the phase difference, and noise can be present in areas with very low signal intensity. To ensure accurate evaluation of the temperature image, a mask is applied before calculating temperature metrics. 

\paragraph{Computation Quantity Metrics.} We evaluate the efficiency of the models by computing their Floating-point Operations per Second (FLOPs), number of parameters (Params), CPU inference time (CPU-T), and effective acceleration ratio ($E_{N=n}$) at a certain undersampling rate, which is calculated using formula \ref{(5)}. It is worth noticing that we place greater emphasis on the effective acceleration ratio, as it can intuitively reflect the acceleration ratio that the model can achieve while considering the model inference time. By combining it with the model's performance for comparison, we can more effectively assess the cost-effectiveness of each model. Additionally, we estimated the total acquisition time (Cost-$N\times$) for magnetic resonance imaging using the $T_{TR}\cdot num_{pe} / N + $ CPU-T formula. As the Fourier inverse transform and PRF temperature measurement have extremely short processing times, they were not included in the time calculation.
\subsubsection{Training Computation}
We employed a mask like the one used in FastMRI to simulate the undersampling process. Specifically, we fully adopted the low-frequency part and uniformly under-sampled the other high-frequency parts, and the proportion of the fully adopted low-frequency part was set to 15\%. We used the AdamW optimizer with a learning rate of 5e-4 and decayed it using a cosine scheduler, and the batch size was set to 8. We conducted experiments separately on both \textit{phantom} and \textit{ex vivo} datasets and trained the models for approximately 200 epochs on an NVIDIA RTX A6000.

\section{Results}
This section presents the experimental results of our study. Firstly, we compare the performance of various deep learning models in the reconstruction of temperature using comparative experiments. Secondly, through ablation experiments, we validate the effectiveness of our proposed method. In addition, we conduct a comprehensive analysis of a long sequence sample, including both time series and consistency analyses. Finally, we demonstrate the resource utilization of different models by presenting parameters and effective acceleration rates.

\subsection{Comparison Study}
\begin{figure*}[ht]
\centerline{\includegraphics[width = 1.\textwidth]{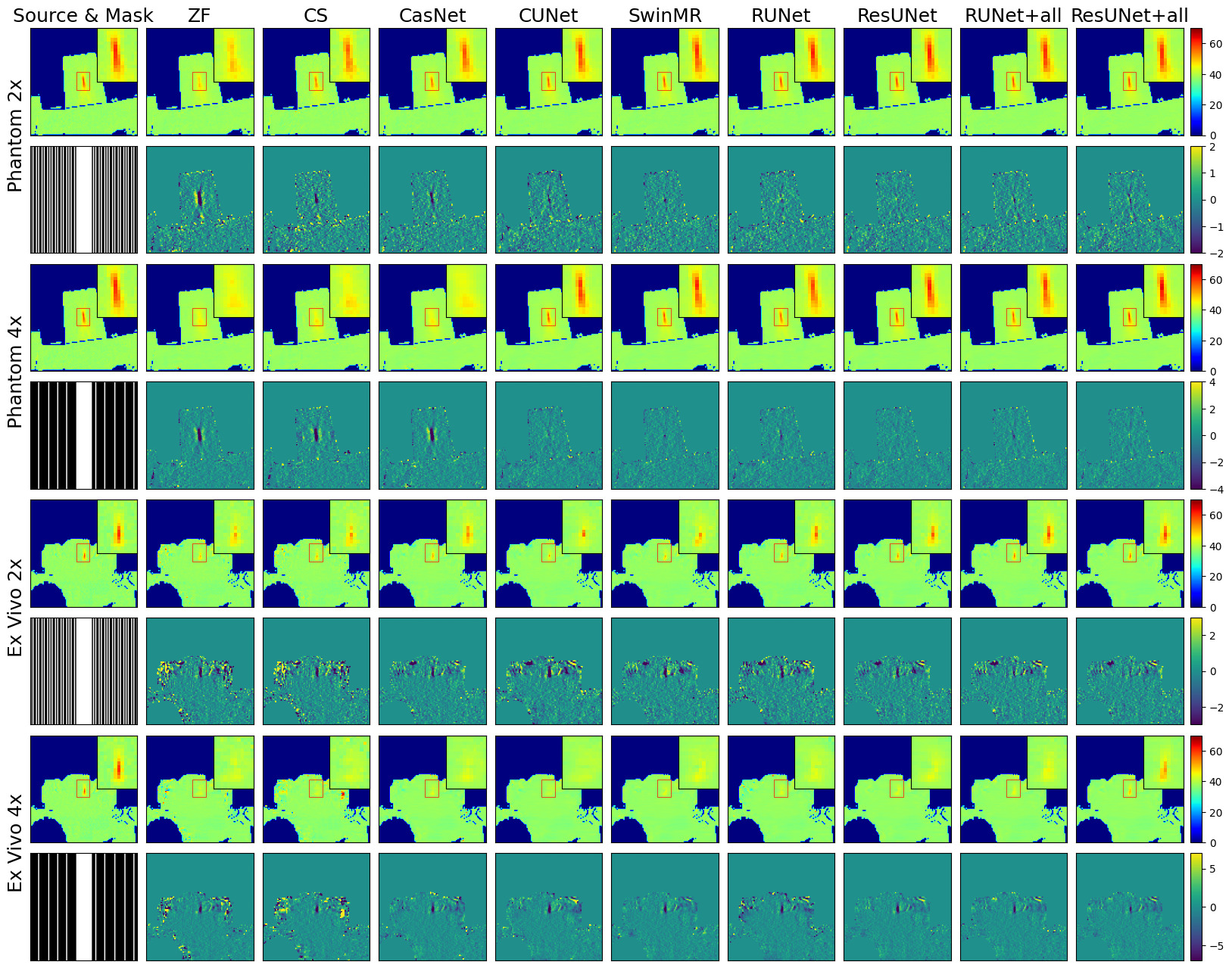}}
\caption{Reconstruction results of different methods for \textit{phantom} and \textit{ex vivo} datasets under $2\times$ and $4\times$ undersampling rates. Each row presents temperature maps and error maps. The first four rows represent the \textit{phantom} results, while the subsequent four rows depict the \textit{ex vivo} results. Different reconstruction methods are displayed in columns 2–10, including ZF (zero-filling), CS, CasNet, CUNet, SwinMR, RUNet, ResUNet, RUNet+all, and ResUNet+all. The first column shows the source temperature maps and under-sampling masks. We zoom in on the central focus area for easier observation. The results illustrate that the proposed optimizing methods for ResUNet, aimed at reconstructing images from various undersampling rates, yield the most favorable outcomes.}
\label{tmap-compare}
\end{figure*}

The comparison results under $2\times$ and $4\times$ undersampling on both \textit{phantom} and \textit{ex vivo} datasets are presented in Table \ref{tab:compare-all}. In our study, we have included the zero-filling (ZF) and compressive sensing (CS) algorithms for comparative analysis. Zero-filling involves filling the under-sampled k-space region with zeros after undersampling, while the compressive sensing algorithm employed is Total Variation Minimization, which makes 200 iterations. It can be observed from the results that the reconstruction performance of RUNet and ResUNet with our optimized methods are superior to that of other methods. In addition, we present the temperature map reconstruction results on specific example samples in Figure \ref{tmap-compare}. It can be observed that the sample using ResUNet+all retains more temperature information in the reconstructed sample, particularly in the \textit{ex vivo} $4\times$ case, where it can display the heated focal temperature, a feature does not present in the results obtained from other methods.

\begin{table*}[htbp]
  \centering
  \caption{Temperature error metrics of different reconstruction methods on \textit{phantom} and \textit{ex vivo} data sets with different undersampling rates}
    \begin{tabular}{cc|cccc|cccc}
    \toprule
          &       & \multicolumn{4}{c|}{\textbf{\textit{phantom}}} & \multicolumn{4}{c}{\textbf{\textit{ex vivo}}} \\
    \textbf{N} & \textbf{Net} & \textbf{$T_{err}$} & \textbf{DICE} & \textbf{STD} & \textbf{RMSE} & \textbf{$T_{err}$} & \textbf{DICE} & \textbf{STD} & \textbf{RMSE} \\
    \midrule
    \multirow{9}[2]{*}{$2\times$} & zf    & 0.310  & 0.402 & 1.219 & 1.286 & 0.542 & 0.507 & 2.403 & 2.751 \\
          & CS    & 0.310 & 0.597 & 1.182 & 1.279 & 0.498 & 0.524 & 2.092 & 2.493 \\
          & CasNet & 0.291 & 0.424 & 1.194 & 1.276 & 0.550  & 0.48  & 2.323 & 2.685 \\
          & CUNet & 0.304 & 0.576 & 1.086 & 1.120  & 0.531 & 0.446 & 2.397 & 2.753 \\
          & SwinMR & 0.259 & 0.759 & 0.935 & 0.950  & 0.523 & 0.519 & 2.218 & 2.566 \\
          & RUNet & 0.302 & 0.669 & 1.055 & 1.080  & 0.559 & 0.44  & 2.505 & 2.876 \\
          & ResUNet & 0.251 & 0.732 & 0.923 & 0.937 & 0.482 & 0.524 & 2.058 & 2.364 \\
          & \cellcolor[rgb]{ .949,  .949,  .949}\textbf{RUNet+all} & \cellcolor[rgb]{ .949,  .949,  .949}\textbf{0.258} & \cellcolor[rgb]{ .949,  .949,  .949}\textbf{0.779} & \cellcolor[rgb]{ .949,  .949,  .949}\textbf{0.903} & \cellcolor[rgb]{ .949,  .949,  .949}\textbf{0.915} & \cellcolor[rgb]{ .949,  .949,  .949}\textbf{0.448} & \cellcolor[rgb]{ .949,  .949,  .949}\textbf{0.537} & \cellcolor[rgb]{ .949,  .949,  .949}\textbf{1.909} & \cellcolor[rgb]{ .949,  .949,  .949}\textbf{2.163} \\
          & \cellcolor[rgb]{ .949,  .949,  .949}\textbf{ResUNet+all} & \cellcolor[rgb]{ .949,  .949,  .949}\textbf{0.245} & \cellcolor[rgb]{ .949,  .949,  .949}\textbf{0.809} & \cellcolor[rgb]{ .949,  .949,  .949}\textbf{0.877} & \cellcolor[rgb]{ .949,  .949,  .949}\textbf{0.888} & \cellcolor[rgb]{ .949,  .949,  .949}\textbf{0.429} & \cellcolor[rgb]{ .949,  .949,  .949}\textbf{0.567} & \cellcolor[rgb]{ .949,  .949,  .949}\textbf{1.802} & \cellcolor[rgb]{ .949,  .949,  .949}\textbf{2.045} \\
    \midrule
    \multirow{9}[2]{*}{$4\times$} & zf    & 0.383 & 0.082 & 1.557 & 1.646 & 0.740  & 0.335 & 3.261 & 3.679 \\
          & CS    & 0.398 & 0.168 & 1.635 & 1.788 & 0.703 & 0.336 & 3.004 & 3.558 \\
          & CasNet & 0.341 & 0.132 & 1.514 & 1.586 & 0.699 & 0.328 & 3.012 & 3.419 \\
          & CUNet & 0.372 & 0.221 & 1.443 & 1.511 & 0.770  & 0.279 & 3.284 & 3.690 \\
          & SwinMR & 0.322 & 0.466 & 1.291 & 1.338 & 0.754 & 0.312 & 3.378 & 3.820 \\
          & RUNet & \textcolor[rgb]{ .212,  .227,  .239}{0.376} & \textcolor[rgb]{ .212,  .227,  .239}{0.256} & \textcolor[rgb]{ .212,  .227,  .239}{1.484} & \textcolor[rgb]{ .212,  .227,  .239}{1.554} & 0.824 & 0.272 & 3.604 & 4.073 \\
          & ResUNet & 0.319 & 0.459 & 1.272 & 1.312 & 0.712 & 0.329 & 3.078 & 3.520 \\
          & \cellcolor[rgb]{ .949,  .949,  .949}\textbf{RUNet+all} & \cellcolor[rgb]{ .949,  .949,  .949}\textbf{0.321} & \cellcolor[rgb]{ .949,  .949,  .949}\textbf{0.488} & \cellcolor[rgb]{ .949,  .949,  .949}\textbf{1.217} & \cellcolor[rgb]{ .949,  .949,  .949}\textbf{1.242} & \cellcolor[rgb]{ .949,  .949,  .949}\textbf{0.635} & \cellcolor[rgb]{ .949,  .949,  .949}\textbf{0.345} & \cellcolor[rgb]{ .949,  .949,  .949}\textbf{2.815} & \cellcolor[rgb]{ .949,  .949,  .949}\textbf{3.167} \\
          & \cellcolor[rgb]{ .949,  .949,  .949}\textbf{ResUNet+all} & \cellcolor[rgb]{ .949,  .949,  .949}\textbf{0.301} & \cellcolor[rgb]{ .949,  .949,  .949}\textbf{0.653} & \cellcolor[rgb]{ .949,  .949,  .949}\textbf{1.126} & \cellcolor[rgb]{ .949,  .949,  .949}\textbf{1.145} & \cellcolor[rgb]{ .949,  .949,  .949}\textbf{0.610} & \cellcolor[rgb]{ .949,  .949,  .949}\textbf{0.365} & \cellcolor[rgb]{ .949,  .949,  .949}\textbf{2.674} & \cellcolor[rgb]{ .949,  .949,  .949}\textbf{3.033} \\
    \bottomrule
    \end{tabular}%
  \label{tab:compare-all}%
\end{table*}%

\begin{figure}[ht]
\centerline{\includegraphics[width = 0.7\textwidth]{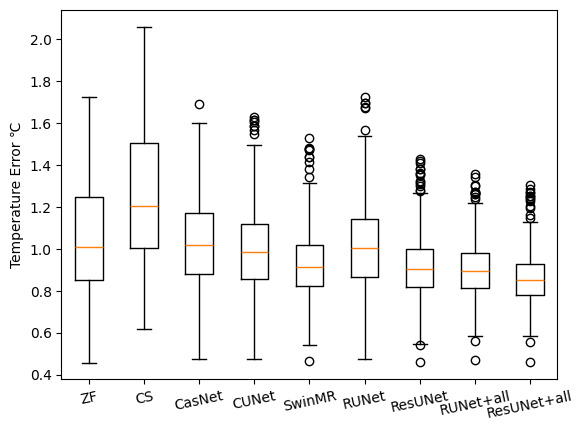}}
\caption{The boxplot shows the average temperature error for each of the nine methods on the \textit{phantom} test set with $4\times$ undersampling. The y-axis represents the temperature error in degrees Celsius. The box represents the interquartile range (IQR), with the median represented by the horizontal line within the box. The whiskers extend to the minimum and maximum values within 1.5 times the IQR, while any outliers are represented as individual points. The results show that the proposed methods outperform the ZF method, with the ResUNet+all method exhibiting the lowest average temperature error. }
\label{boxplot}
\end{figure}

To visually compare the efficacy of various methods for rapid temperature measurement, we calculated the mean temperature error using pixels above 43 ºC within the temperature map patches reconstructed by each model. As an example, we utilized the \textit{phantom} sub-dataset with $4\times$ undersampling. Subsequently, we generated box plots for all samples in the test set, as depicted in Figure \ref{boxplot}. The results indicate that the proposed methods are superior to the ZF method, with the ResUNet+all method demonstrating the lowest average temperature error.

\subsection{Ablation Study}

\begin{figure}[t]
\centerline{\includegraphics[width = 0.7\textwidth]{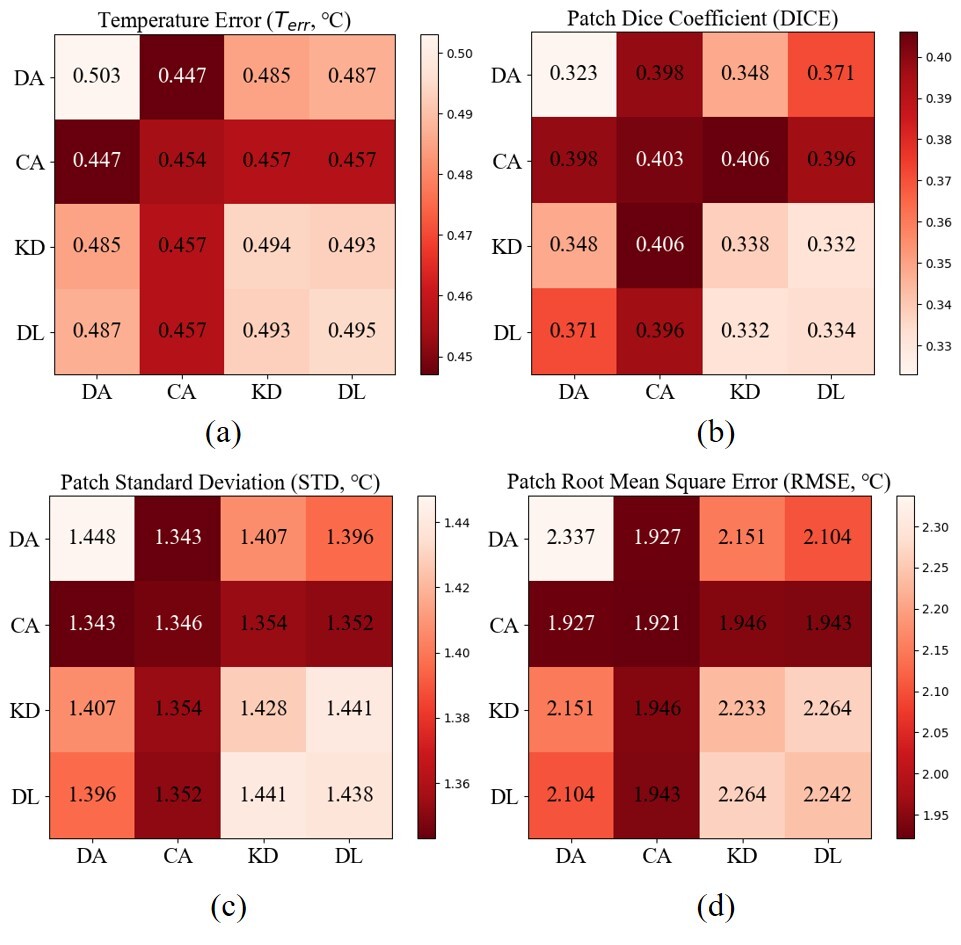}}
\caption{Heat maps of temperature metrics for four reconstruction modules. This figure shows the heat maps of four temperature metrics obtained from pairwise combinations of four reconstruction modules. For each temperature metric, the four modules were pairwise combined to generate 16 combinations, which were evaluated to generate a heat map. The heat maps visually demonstrate the impact of different module combinations on temperature metrics. The values in the heat maps (a), (b), (c), and (d) represent the magnitude of the different temperature metrics; darker colors indicate better reconstruction effects.}
\label{heat_maps}
\end{figure}

To showcase the efficacy of our method, we conducted a series of ablation experiments on the RUNet model using \textit{phantom} validation datasets that were subjected to $4\times$ undersampling. As shown in Table \ref{tab:ablation-all}, we incorporated four individual modules (DA for diffusion model augmentation, CA for complex-valued data augmentation, KD for knowledge distillation, and DL for decoupled loss) into the baseline model across all four temperature metrics, resulting in varying degrees of improvement. In addition, we combined the four modules in all possible combinations and generated a $4\times4$ heat map matrix to visualize the resulting temperature indicators; different combinations of the modules have varying effects on specific indicators, as illustrated in Figure \ref{heat_maps}. However, as an overall observation, it can be inferred that complex-valued data augmentation had the most substantial contribution.

\begin{table*}[b]
  \centering
  \caption{Temperature error metrics of RUNet with different modules on $4\times$ undersampled \textit{phantom} dataset}
    \begin{tabular}{cccc|cccc|cccc}
    \toprule
    \multicolumn{4}{c|}{\textbf{Modules}} & \multicolumn{4}{c}{\textbf{\textit{phantom}}} & \multicolumn{4}{c}{\textbf{\textit{ex vivo}}} \\
    \textbf{DA} & \textbf{CA} & \textbf{KD} & \textbf{DL} & \textbf{$T_{err}$} & \textbf{DICE} & \textbf{STD} & \textbf{RMSE} & \textbf{$T_{err}$} & \textbf{DICE} & \textbf{STD} & \textbf{RMSE} \\
    \midrule
          &       &       &       & \textcolor[rgb]{ .651,  .651,  .651}{0.505} & \textcolor[rgb]{ .651,  .651,  .651}{0.327} & \textcolor[rgb]{ .651,  .651,  .651}{1.473} & \textcolor[rgb]{ .651,  .651,  .651}{1.538} & \textcolor[rgb]{ .651,  .651,  .651}{0.822} & \textcolor[rgb]{ .651,  .651,  .651}{0.266} & \textcolor[rgb]{ .651,  .651,  .651}{2.750} & \textcolor[rgb]{ .651,  .651,  .651}{2.971} \\
    \checkmark     &       &       &       & 0.503 & 0.323 & 1.448 & 1.529 & 0.749 & 0.231 & 2.575 & 2.788 \\
          & \checkmark     &       &       & 0.454 & 0.403 & 1.346 & 1.386 & 0.612 & 0.311 & 2.248 & 2.454 \\
          &       & \checkmark     &       & 0.494 & 0.338 & 1.428 & 1.494 & 0.775 & 0.266 & 2.727 & 2.974 \\
          &       &       & \checkmark     & 0.495 & 0.334 & 1.438 & 1.497 & 0.793 & 0.224 & 2.692 & 2.892 \\
    \rowcolor[rgb]{ .949,  .949,  .949} \checkmark     & \checkmark     & \checkmark     & \checkmark     & \textbf{0.447} & \textbf{0.415} & \textbf{1.337} & \textbf{1.382} & \textbf{0.594} & \textbf{0.318} & \textbf{2.227} & \textbf{2.420} \\
    \bottomrule
    \end{tabular}%
  \label{tab:ablation-all}%
\end{table*}%

\subsection{Time-Consuming Study}

To evaluate the resource utilization of each network we used four key indicators: number of parameters (Params), number of floating-point operations (FLOPs), CPU inference time (CPU-T), and total cost and the effective acceleration rate under $2\times$ (Cost-$2\times$, $E_{N=2}$) and $4\times$ (Cost-$4\times$, $E_{N=4}$) undersampling. Here, the performance of CPU-T is evaluated by conducting 1000 forward processing runs and calculating the average processing time on an Intel(R) Xeon(R) Gold 6248R CPU. These indicators were selected to provide a comprehensive assessment of the network's resource utilization in terms of its computational complexity, memory usage, and inference speed. Through our evaluation of these indicators, we were able to gain insights into the efficiency and effectiveness of each network.

Our evaluation results are presented in Table \ref{tab:cost}. Compared with SwinMR and CS methods, RUNet and ResUNet exhibit the shortest CPU running time and the highest effective acceleration rate among all the evaluated networks. These findings suggest that RUNet and ResUNet may be particularly well-suited for resource constrained applications that require low-latency and high-throughput processing. Furthermore, in conjunction with our evaluation of temperature indicators, our results show that the addition of the ResUNet+all network yields a remarkably high cost-effectiveness ratio. This suggests that the proposed optimizing modules may serve as a valuable augmentation technique for improving the performance and efficiency of the UNet network, especially in applications where resource utilization is a critical consideration.

\begin{table*}[b]
  \centering
  \caption{Resource utilization metrics of different models under different under-sampling rates}
    \begin{tabular}{c|ccccccc}
    \toprule
    \multirow{2}[2]{*}{\textbf{ Methods}} & \multicolumn{7}{c}{\textbf{Metrics}} \\
          & \textbf{Params(M)} & \textbf{FLOPs(G)} & \textbf{CPU-T(s)} & \textbf{Cost-$2\times$(s)} & \textbf{$E_{N=2}$} & \textbf{Cost-$4\times$(s)} & \textbf{$E_{N=4}$} \\
    \midrule
    \textcolor[rgb]{ .651,  .651,  .651}{ZF} & \textcolor[rgb]{ .651,  .651,  .651}{\textbf{-}} & \textcolor[rgb]{ .651,  .651,  .651}{\textbf{-}} & \textcolor[rgb]{ .651,  .651,  .651}{\textbf{-}} & \textcolor[rgb]{ .651,  .651,  .651}{0.768} & \textcolor[rgb]{ .651,  .651,  .651}{2.0} & \textcolor[rgb]{ .651,  .651,  .651}{0.384} & \textcolor[rgb]{ .651,  .651,  .651}{4.0} \\
    CS    & - & - & 22.653 & 23.421 & 0.1   & 23.037 & 0.0 \\
    CasNet & \textbf{0.10} & 4.36  & 0.0431 & 0.811 & 1.9   & 0.427 & 3.6 \\
    CUNet & 3.87  & \textbf{1.68} & 0.0976 & 0.866 & 1.8   & 0.482 & 3.2 \\
    SwinMR & 11.45 & 105.74 & 0.4370 & 1.205 & 1.3   & 0.821 & 1.9 \\
    RUNet-Tea & 123.70 & 26.67 & 0.1112 & 0.879 & 1.7   & 0.495 & 3.1 \\
    ResUNet-Tea & 32.83 & 51.61 & 0.0687 & 0.837 & 1.8   & 0.453 & 3.4 \\
    \rowcolor[rgb]{ .949,  .949,  .949} RUNet & 7.74  & 1.69  & \textbf{0.0282} & \textbf{0.796} & \textbf{1.9} & \textbf{0.412} & \textbf{3.7} \\
    \rowcolor[rgb]{ .949,  .949,  .949} ResUNet & 2.06  & 3.23  & \textbf{0.0277} & \textbf{0.796} & \textbf{1.9} & \textbf{0.412} & \textbf{3.7} \\
    \bottomrule
    \end{tabular}%
  \label{tab:cost}%
\end{table*}%

\subsection{Long sequence sample study}

Applying rapid temperature measurement to practical devices improves the temporal resolution of the temperature measurement process. This improvement is reflected in an increased number of frames when the HIFU heating power and ablation duration remain constant. To examine the potential impact of changes in temporal resolution on temperature measurement, we analyzed temperature image samples from simulated long-sequence data of a \textit{phantom} model with improved temporal resolution.

In this study we selected 32-frame long-sequence samples from the test set of the \textit{phantom} model and extracted every other frame to simulate a sequence of 16 frames representing full sampling conditions for a given heating duration. The remaining 32 frames simulated a sequence obtained under $2\times$ undersampling. Using ResUNet+all, we recorded temperatures of the $3\times3$ pixel block at the center of the original reconstructed temperature images. 

\begin{figure}[h]
\centerline{\includegraphics[width = 1.\textwidth]{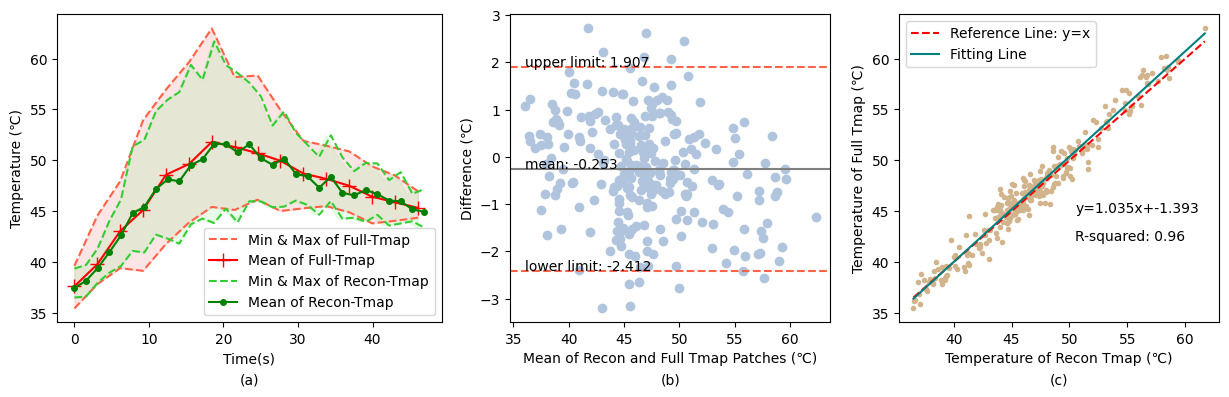}}
\caption{Evaluation results of ResUNet+all method on a long sequence of \textit{phantom} dataset test sample. The sequence contains 32 frames with $2\times$ undersampling, and the evaluation focuses on the $3\times3$-pixel block at the center of the heating point, which contains the most valuable temperature information. (a) Time-Temperature Plot: This subplot shows the time-temperature curve of the original and reconstructed temperature maps. The shaded area represents the range of maximum and minimum temperatures in the $3\times3$-pixel block. (b) Bland-Altman Plot: The horizontal and vertical coordinates of the subgraph represent the mean temperature and temperature error in the  $3\times3$-pixel blocks of the original temperature map and the reconstructed temperature map, respectively. The solid gray line represents the mean of the temperature error, and the dotted red line represents the 95\% confidence interval. (c) The horizontal and vertical coordinates represent the pixel block temperature in the reconstructed and original temperature maps, respectively. The red dotted line represents perfect alignment ($y=x$), while the blue solid line represents the fitted correlation. (slope/intercept = $1.035/1.393$, $R^2 = 0.96$, $bias = 1.39$ ℃, 95\% CI: $-0.253\pm2.160$ ℃)}
\label{long_seq}
\end{figure}

Temperature-time curves were plotted and analyzed using Bland-Altman and linear regression methods. The calculation of time on the horizontal axis followed the same method as Cost-$N\times$. Additionally, we incorporated the inference time of the model into the plotting of the temperature-time curves, where a longer inference time resulted in a rightward shift compared to the fully sampled temperature curve. A larger shift indicated a greater impact on temporal resolution improvement, reflecting a poor performance, while a smaller shift indicated a better performance.

The results, presented in Figure \ref{long_seq}, demonstrate that the reconstructed focus closely aligns with the temperature-time curve of the fully sampled images. There is no significant deviation observed in the curve, suggesting that the model exhibits strong real-time inference capability. Most of the data points fall within the 95\% confidence interval, with upper and lower limits within $\pm3$ ℃, indicating a strong linear relationship between the reconstructed and fully-sampled temperature maps, suggesting that the reconstructed temperature map is highly consistent with the fully-sampled temperature map.

\section{Discussion}

In this article, we present an introduction to the application of deep learning methods for rapid magnetic resonance thermometry. Previously, the fast readout patterns have been studied for temperature measurements, such as spiral and radial strategies, EPI sequence\cite{gaurAcceleratedMRIThermometry2015,kim2023motion,odeen2014sampling}. The deep learning-based rapid reconstruction here was appropriate for them. In the case of under-sampling, these sequences can achieve a higher temporal resolution, combining our proposed methods. Theoretically, our approach can be applied by retraining whenever the acceleration is realized via undersampling. Specifically, we focus on enhancing the effective acceleration rate and improving model performance within a short inference time. To achieve this, we propose a series of model-agnostic techniques and validate the effectiveness of each module through experimental verification. The assembly of under-sampling and deep learning reconstruction for fast temperature measurements has quite a few benefits. For example, Undersampling means fewer phase encoding and less B0 field drift\cite{wang2018method}. Without considering the loss of signal-to-noise ratio (SNR), the measured temperature should be more accurate. On the subject of motion-induced artifacts, our proposed deep-learning reconstruction module can also be improved to make it insensitive to respiration and other movement by modifying the neural network module. Once the fast thermometry can be realized, the volumetric temperature monitoring covering the whole focal area will be easier.

In our experiments, we employed only a maximum undersampling rate of $4\times$, instead of the up to $10\times$ rate used by fastMRI. This choice was influenced by the limitations of the image resolution and signal-to-noise ratio in our acquired images. The number of phase encodings in k-space was also relatively low, and the signal was not highly concentrated at the center. These factors resulted in significant signal loss when using excessively high undersampling rates, making reconstruction difficult. Interestingly, the compromised resolution and signal-to-noise ratio in the acquired dataset are flaws due to equipment design, which prioritized faster temperature mapping at the expense of spatial resolution; this highlights the importance of rapid thermometry. With the same temperature mapping speed, higher spatial resolution and signal-to-noise ratio can be achieved, leading to higher-quality images. Consequently, these higher-quality images can then be subjected to higher undersampling rates.

To address this, we conducted an experiment using a phantom dataset and performed interpolation to simulate the acquisition of high-resolution images. Subsequently, we applied undersampling rates of $6\times$, $8\times$ and $10\times$ to these images and compared the results with those obtained from the original resolution images. The obtained RMSE values for the temperature map patches were 0.610℃, 0.704℃, and 0.724℃, respectively. These values closely align with the results obtained for $2\times$ and $4\times$ resolutions in a $96\times96$ format.

Furthermore, our study only utilized \textit{phantom} and \textit{ex vivo} tissue datasets. The temperature distribution in actual human tissue is more complex. Therefore, in the future, we plan to conduct testing and research on live animal models and specific human tissue datasets to further investigate and validate our findings.

\section{Conclusion}

This paper presents the first formal investigation into the application of deep learning methods for magnetic resonance temperature measurement. To the best of our knowledge, this is the first comprehensive study to explore the use of deep learning methods for this purpose. We have made our code publicly available and have proposed the use of four optimizing modules to enhance model performance without increasing the number of parameters or computational complexity. We compared various existing MR re-construction models and demonstrated the effectiveness of our proposed method, as well as its resource-saving characteristics. We hope that our research will serve as inspiration for further investigations related to MRI temperature measurement. Moving forward, we plan to explore end-to-end approaches that incorporate temporal information, as well as investigate the feasibility of adopting reference-free imaging techniques for rapid temperature measurement.


\bibliographystyle{unsrt}  
\bibliography{references}  






\end{document}